\documentclass[twocolumn]{aastex631}

\usepackage{amsmath}
\usepackage{graphicx}
\usepackage{dcolumn}
\usepackage{bm}
\usepackage{xcolor}

\begin{document}

\title{Particle acceleration in relativistic Alfv\'enic turbulence}

\correspondingauthor{Cristian Vega}
\email{csvega@wisc.edu}

\author{Cristian Vega}
\affiliation{Department of Physics, University of Wisconsin at Madison, Madison, Wisconsin 53706, USA}

\author[0000-0001-6252-5169]{Stanislav Boldyrev}
\affiliation{Department of Physics, University of Wisconsin at Madison, Madison, Wisconsin 53706, USA}
\affiliation{Center for Space Plasma Physics, Space Science Institute, Boulder, Colorado 80301, USA}

\author[0000-0003-1745-7587]{Vadim Roytershteyn}
\affiliation{Center for Space Plasma Physics, Space Science Institute, Boulder, Colorado 80301, USA}



\begin{abstract}
Strong magnetically dominated Alfv\'enic turbulence is an efficient engine of non-thermal particle acceleration in a relativistic collisionless plasma. We argue that in the limit of strong magnetization, the type of energy distribution attained by accelerated particles depends on the relative strengths of turbulent fluctuations $\delta B_0$ and the guide field $B_0$. If $\delta B_0\ll B_0$, the particle magnetic moments are conserved and the acceleration is provided by magnetic curvature drifts. Curvature acceleration energizes particles in the direction parallel to the magnetic field lines, resulting in log-normal tails of particle energy distribution functions. Conversely, if $\delta B_0 \gtrsim B_0$, interactions of energetic particles with intense turbulent structures can scatter particles, creating a population with large pitch angles. In this case, magnetic mirror effects become important, and turbulent acceleration leads to power-law tails of the energy distribution functions.
~
\end{abstract}



\section{Introduction} \label{sec:intro}
Turbulence governed by Alfv\'enic fluctuations is ubiquitous in magnetized space and astrophysical plasmas. In many natural systems, turbulent fluctuations exist in a broad range of scales, from the macroscopic size of the system to the micro scales associated with particle inertial or gyroscales. When the collisions between the particles are less frequent than the typical turbulent interactions, the particles can be energized by turbulence in a non-thermal fashion, so that their momentum distribution function significantly deviates from a Maxwellian. A small fraction of the particles can even be accelerated to form a runaway population, leading to a power-law tail of the energy distribution function \cite[e.g.,][]{zhdankin2017a,comisso2019,zhdankin2020,nattila2020,nattila2022,demidem2020,trotta2020,pezzi2022,vega2022a,bresci2022,comisso2022,vega2023}. 

Non-thermal energetic particles are essential components of many natural and laboratory plasmas. They may influence plasma dynamics and define the spectra of radiation coming from astrophysical objects.  Particle acceleration plays a fundamental role in the solar flares, radiation coming from pulsar wind nebulae, jets from active galactic nuclei, non-thermal Galactic radio filaments, gamma-ray bursts, and other astrophysical phenomena \cite[e.g.,][]{bykov1996,selkowitz2004,petrosian2004,tramacere2011,bian2012,xu2019,asano2020,yusef-zadeh2022}.

It is well established that efficient particle acceleration can be provided by magnetic plasma structures such as collisionless shocks \cite[e.g.,][]{blandford1987,marcowith2016} or intense reconnecting current sheets \cite[e.g.,][]{uzdensky2011,drake2013,sironi2014,sironi2022,french2023,guo2020,guo2023}. Astrophysical Alfv\'enic turbulence contains a hierarchy of strongly nonlinear fluctuations (eddies) that typically span an enormous range of spatial and temporal scales. Turbulent flows also spontaneously generate intermittent structures such as magnetic discontinuities, shocks, vortices, and reconnecting current layers \cite[e.g.,][]{matthaeus_turbulent_1986,loureiro2017,loureiro2018,loureiro2019,roytershteyn19,walker2018,mallet2017a,comisso2019,boldyrev_2017,boldyrev_loureiro2018,boldyrev2019,vega2020,dong2022}. Although numerical simulations suggest that strong Alfv\'enic turbulence is an efficient engine for non-thermal particle acceleration, at present, there is no complete understanding of this process. Recent analytical and numerical studies, however, indicate that spontaneously generated turbulent structures may indeed play an essential role in particle energization \cite[e.g.,][]{trotta2020,ergun2020a,lemoine2021,pezzi2022,bresci2022,vega2022a,xu2023,lemoine2023}.

It is reasonable to believe that particle energization by magnetic fluctuations is related to the properties of Alfv\'enic turbulence itself. 
Let us assume that turbulence is generated by forces or instabilities operating at some scale $L_0$ that is much larger than the plasma microscales. We will refer to this scale as the outer scale of turbulence. We denote the strength of magnetic fluctuations at the outer scale as $\delta B_0$ and also assume that the system is immersed in a uniform background magnetic field $B_0$. The presence of the background field is important only if $B_0$ is greater than $\delta B_0$, which may happen when the background field is imposed by external mechanisms not related to considered turbulence. Examples are magnetospheres of planets and stars or external coils in laboratory experiments. Conversely, when the uniform field is weak, $B_0 \ll \delta B_0$, its presence is not significant for the statistics of turbulent fluctuations.  

For definiteness, assume that we are dealing with a pair plasma. We consider the limit of magnetically dominated turbulence characterized by a large magnetization parameter, 
\begin{eqnarray}
{\tilde \sigma}=\frac{(\delta B_0)^2}{4\pi n_0 w_0 m_e c^2} \gtrsim 1,\label{sigma}
\end{eqnarray}
where $w_0$ is the specific enthalpy associated with the distribution of bulk plasma electrons (say, thermal distribution), and $n_0$ is their density.  When ${\tilde \sigma}\gg 1$, turbulent fluctuations are intense enough to heat plasma to ultra-relativistic energies, $w_0\gg 1$. Supra-thermal particles accelerated by such turbulence will also have ultra-relativistic velocities.  Since the particle energy is given by $\gamma m_ec^2$, we will discuss particle acceleration by considering the evolution of the Lorenz factor~$\gamma$.

In the presence of a strong background magnetic field, the correlation scales of turbulence are different in the directions parallel and perpendicular to this field.\footnote{As discussed previously \cite[e.g.,][]{zhdankin2018c,nattila2020,vega2022a} and also mentioned in section~\ref{accel_section}, relativistic plasma turbulence rapidly relaxes to the state with ultra-relativistic particle temperature but mildly relativistic bulk fluctuations. Therefore, turbulent fluctuations at large hydrodynamic scales can be described, to the leading order, in the framework of non-relativistic MHD.} We denote the field-parallel outer scale as $L_{\|, 0}$ and the corresponding field-perpendicular scale $L_{\perp,0}$. These scales are related by the so-called critical balance condition, $L_{\perp,0}/L_{\|,0}\sim \delta B_0/B_0$ \cite[][]{goldreich_toward_1995}. 
In a turbulent state, the strength of Alfv\'enic fluctuations decreases with the field-parallel scale as $\delta B_{\perp}/\delta B_0 \sim (l/L_{\|,0})^{1/2}$, and with the field-perpendicular scale approximately as $\delta B_{\perp}/\delta B_0 \sim (\lambda/L_{\perp,0})^{\alpha}$, with $\alpha \approx 1/3$ or $\alpha \approx 1/4$, depending on the phenomenological model \cite[e.g.,][]{goldreich_toward_1995,boldyrev2006}. The exact value of the exponent $\alpha$ is not important for our consideration.  

Whether a uniform background magnetic field is imposed or not, turbulent subdomains of scales $\lambda\ll L_{\perp,0}$ possess a nearly uniform background magnetic field provided by the outer-scale eddies. Such subdomains may however be separated by relatively sharp boundaries, with thickness as small as the inner scale of turbulence \cite[e.g.,][]{borovsky2008,zhdankin2012a,zhdankin2012b}. In the case of a pair plasma, it is the electron inertial scale. 
The strongest magnetic-field variations, $\delta {\bm B}_0$, occur across the boundaries between the largest eddies. Such spatially intermittent structures occupy a small volume, so they do not significantly affect the Fourier spectra of turbulent fluctuations. However, they may contribute to particle heating and acceleration \cite[e.g.,][]{zhdankin2016,zhdankin2017}. 

In this work, we propose a description of particle acceleration caused by their interactions with a hierarchy of nonlinear fluctuations (eddies) produced by Alfv\'enic turbulence.  In a collisionless plasma, turbulence can be set up at outer scales in a variety of different ways, which may affect the processes of particle heating and acceleration. 
In our discussion, we assume the following ``self-consistent" way of turbulence excitation. We assume that the initial particle distribution is mildly relativistic and isotropic, and that the turbulence is driven by initially strong magnetic perturbations, ${\tilde \sigma}_0\gg 1$,  that are allowed to decay freely. In this setting, the details of the initial particle distribution are not essential; rather, both the bulk and non-thermal parts of the energy distribution function are self-consistently shaped by turbulence. We also neglect particle cooling by radiation.

{We argue that in the considered limit of strong magnetization, the nature of the resulting particle acceleration process is governed by the relative strength of the guide field. We start with the case of a strong guide field, $B_0\gg \delta B_0$. In this case, decaying Alfv\'enic turbulence heats the particles along the magnetic field lines, which leads to a nearly one-dimensional ultra-relativistic particle distribution function. (Such strongly anisotropic distributions are also common in astrophysical systems where the particles are significantly cooled in the field-perpendicular direction by synchrotron radiation.) We argue that in the case of a strong guide field, the acceleration is relatively inefficient. In this limit, particle magnetic moments are preserved, particles are accelerated in the direction parallel to the magnetic field, and their energy distributions generally have log-normal statistics. We then turn to the limit of a moderate guide field, $ B_0\lesssim \delta B_0$. In this case, magnetic moments of energetic particles are not conserved during interactions with intense turbulent structures, the acceleration process becomes more efficient, and the particle energy distribution functions develop power-law tails. }


\section{Particle acceleration by Alfv\'enic turbulence}
\label{accel_section}
A relativistic particle whose gyroradius is much smaller than the typical scale of magnetic field variations preserves its first adiabatic invariant, the magnetic moment. Such a particle can be accelerated by turbulent fluctuations in several different ways. These can be divided into three categories. First is the acceleration by a parallel electric field, that is, the electric-field component parallel to the magnetic field. Second is the acceleration due to magnetic curvature drifts. And third is the acceleration by magnetic mirror forces. In all the cases, a particle is, of course, accelerated by an electric field. The proposed division into the three categories helps to associate the acceleration mechanisms with turbulent structures. Below, we analyze these cases concentrating on an  ultra-relativistic electron-positron pair plasma.

\subsection{Acceleration by parallel electric field}
The parallel electric field fluctuations are relatively weak in Alfv\'enic turbulence with a strong guide field. For instance, in a magnetically dominated ultrarelativistic plasma with one-dimensional momentum distribution of particles,\footnote{This is certainly an idealization, since in reality (and in numerical simulations) the particle distributions are not strictly one-dimensional. Moreover, turbulence may contain a small admixture of ordinary and extraordinary modes~\cite[e.g.,][]{vega2024}.} the linear wave analysis in \cite[][]{vega2024} yields for the parallel and perpendicular electric fluctuations at scales $\lambda\gtrsim d_{rel}$:
\begin{eqnarray}
\label{E_par}
\frac{E_{\|}}{E_{\perp}} \approx \frac{1}{w_0^2}\,{k_\|k_\perp d_{rel}^2}, 
\end{eqnarray}
see Appendix~\ref{e_polarization}: Eq.~(\ref{A3}). Here, $k_\perp=2\pi/\lambda$ is the field-perpendicular wave number of the electric field, $k_\|=2\pi/l$ the field-parallel wavenumber, $w_0$ is the enthalpy per particle associated with the (relativistic) distribution function of the bulk electrons, and $d_{rel}$ is the corresponding electron inertial scale. 

The parallel electric field is strongest at the electron inertial scale $k_\perp d_{rel}\approx 1$. As a relativistic electron propagates through such an eddy, its energy gain is
\begin{eqnarray}
m_ec^2\Delta \gamma = q E_\| l\sim 2\pi q\, d_{rel} \,\delta B_\perp(l)/w_0^2,
\end{eqnarray} 
where we used the fact that the electric and magnetic fluctuations are in approximate equipartition in magnetically-dominated turbulence, $E_\perp(l)\sim \delta B_\perp(l)$. 
Substituting here the scaling $\delta B_\perp(l)\sim \delta B_0\left(l/L_\|\right)^{1/2}$, we express the energy gain as
\begin{eqnarray}
\Delta \gamma \sim \frac{\pi\sqrt{2{\tilde \sigma}}}{w_0}\left(\frac{l}{L_\|}\right)^{1/2}. 
\end{eqnarray}

Depending on the direction of the parallel electric field, an electron can either gain or lose energy in an individual interaction. The typical energy gain is then proportional to the square root of the number of electron interactions with the eddies of scale~$l$. We can express this number as $N_0 L_\|/l$, where $N_0$ is the number of outer scale distances crossed by the electron. As a result, we obtain a typical energy gain after $N_0$ large-scale crossing times:
\begin{eqnarray}
\Delta \gamma \sim \frac{\pi\sqrt{2{\tilde \sigma}}}{w_0}\sqrt{N_0}.   
\end{eqnarray}

When ${\tilde \sigma}$ is initially large, which is the case in some numerical setups of decaying turbulence \cite[e.g.,][]{comisso2018,comisso2019,vega2022a,vega2022b,vega2023,vega2024}, the parallel particle heating and acceleration are initially strong. However, as the magnetic perturbations relax and release energy to plasma particles, plasma magnetization ${\tilde \sigma}$ decreases while (in the absence of significant radiative cooling) the particle enthalpy $w_0$ increases. As a result, parallel electric heating by Alfv\'en modes becomes progressively less significant. Moreover, the parallel electric acceleration is linear (or algebraic) in time, as opposed to the exponential acceleration due to curvature and mirror effects discussed below. Finally, we note that parallel heating and acceleration increase the field-parallel momentum of a particle, but not its field-perpendicular momentum. Therefore, the particle's pitch angle, that is, the angle between the particle velocity and the magnetic field line, {\it decreases} as a result of such a process. 

\subsection{Curvature acceleration}
\label{curve_section}
\begin{figure}[tb]
\centering
\includegraphics[width=1.2\columnwidth]{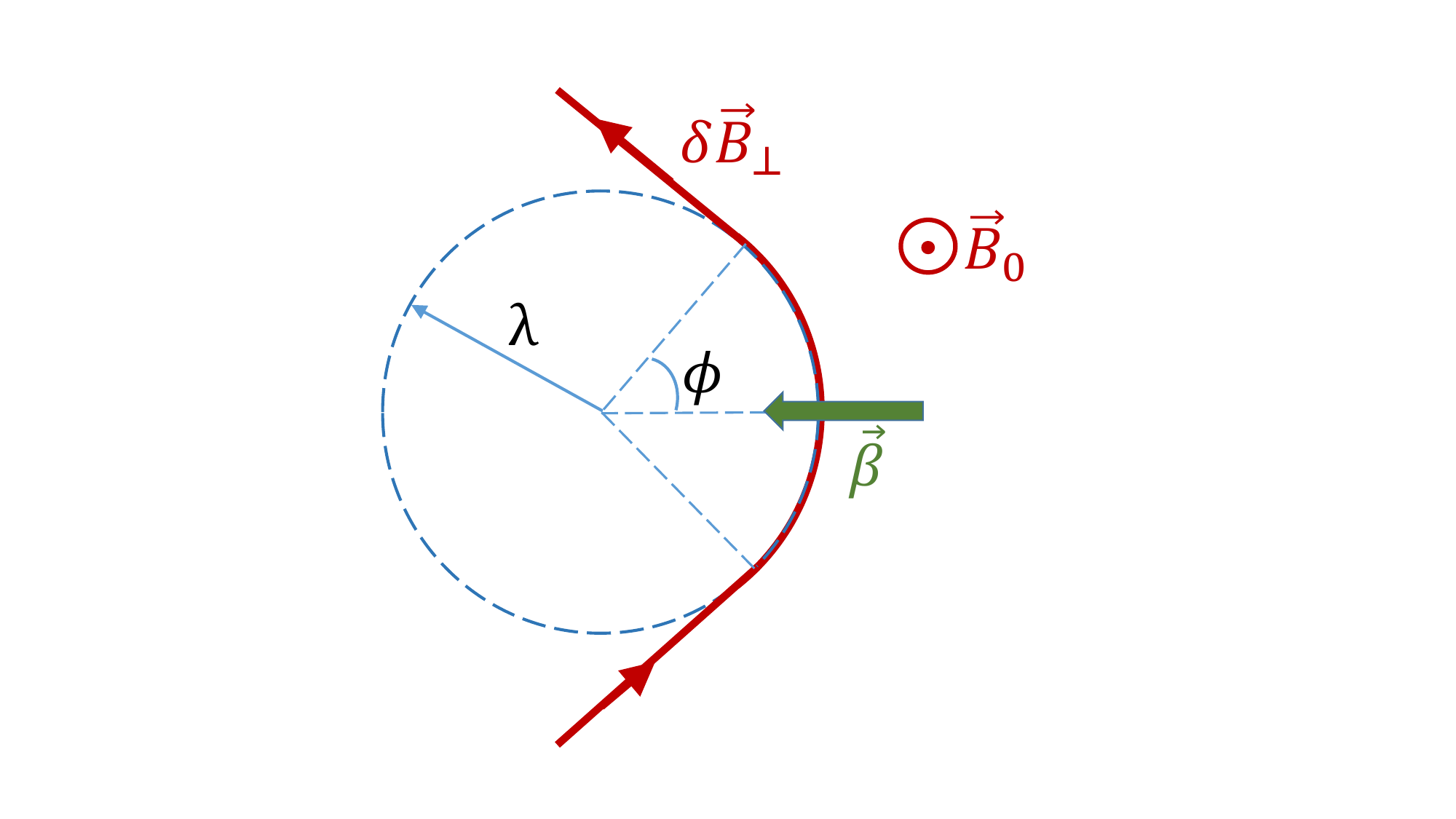}
\caption{Sketch of a curved magnetic field line. The projection of the line on the plane normal to the background magnetic field ${\bm B}_0$ is shown. For simplicity, it is assumed that the magnetic structure moves with velocity ${\bm \beta}$ in the direction parallel to the field-line curvature. An electron propagating along the curved line, experiences curvature acceleration.}
\label{curve_accel}
\end{figure}

When the parallel electric field effects become negligible and the particle magnetic moment is conserved, the dominant acceleration is provided by the curvature drift. Indeed, this drift does not vanish in the limit of small pitch angles and, therefore, it remains efficient when the particle's parallel momentum increases. To discuss the curvature acceleration, assume that the uniform component of the magnetic field is in the $z$-direction. The Alfv\'enic magnetic perturbation $\delta {\bm B}_\perp$ is then in the $x-y$ plane. A sketch of the projection of a curved magnetic-field line onto the $x-y$ plane is given in Fig.~\ref{curve_accel}.  
We assume that such a magnetic structure is moving with the velocity ${\bm \beta}={\bm u}_{E\times B}/c$, as shown. Here, the E-cross-B velocity, ${\bm u}_{E\times B}=c{\bm E}\times {\bm B}/B^2$, describes the bulk (or fluid) velocity fluctuations of the plasma. Such fluid fluctuations are typically only mildly relativistic \cite[e.g.,][]{zhdankin2018c,vega2022a}, so for simplicity, we can neglect relativistic factors associated with the fluid motion. A relativistic electron propagating along the magnetic field line experiences curvature acceleration. When the pitch angle is small, the electron moves along the line at nearly the speed of light. As the electron passes the curved part of the line, its energy changes according to \cite[e.g.,][]{northrop1963}:
\begin{eqnarray}
\label{c_ac}
\Delta \ln(\gamma) \sim {\beta} \frac{S}{R_c},    
\end{eqnarray}
where $R_c$ is the curvature radius of the total magnetic field line (that is, not its projection in Fig.~\ref{curve_accel}) and $S$ is the corresponding length of the curved path.  Here, $\beta ={\bm \beta}\cdot {\bm R}_c/R_c$ is the projection of the velocity ${\bm \beta}$ onto the direction of the curvature vector, that is, the vector connecting the point on the trajectory with the curvature center. For the collision shown in Figure~\ref{curve_accel}, the particle gains energy, $\beta>0$. Such collisions are more probable, so the particle gains energy on average. 

A standard geometric calculation gives for the curvature radius
\begin{eqnarray}
\label{R_c}
R_c=\lambda\left[1+\left(\frac{B_0}{\delta B_\perp}\right)^2\right],
\end{eqnarray}
and for the path length
\begin{eqnarray}
\label{S}
S=2\phi\lambda\left[1+\left(\frac{B_0}{\delta B_\perp}\right)^2\right]^{1/2},    
\end{eqnarray}
where fore an estimate, one can assume~$2\phi\lesssim \pi$. Substituting these expressions into Eq.~(\ref{c_ac}), we obtain
\begin{eqnarray}
\label{delta}
\Delta \ln(\gamma)\sim 2\phi \beta  \left[1+\left(\frac{B_0}{\delta B_\perp}\right)^2\right]^{-1/2}.  
\end{eqnarray}
We note that this result does not depend on the magnitude of the curvature given by the structure's scale,~$\lambda$. Indeed, a smaller curvature radius would provide a stronger acceleration, however, the propagation path would be shorter, resulting in the same energy gain. 

We also note that Eq.~(\ref{S}) relates the field-parallel and field-perpendicular dimensions of the turbulent eddy considered in Fig.~(\ref{curve_accel}). Consider small enough scales where $\delta B_\perp/B_0 \ll 1$. If we formally introduce the field-parallel and field-perpendicular wavenumbers as $k_\perp=2\pi/(2\lambda)$ and $k_\|=2\pi/S$, and set $2\phi=\pi$, we derive
\begin{eqnarray}
\frac{k_\perp}{k_\|}\frac{\delta B_\perp}{B_0}=\frac{\pi}{2}.    
\end{eqnarray}
This formula is the critical balance condition of \citet{goldreich_toward_1995} that takes into account the magnetic-field-line curvature effects.

The analysis of particle acceleration can be advanced further if the magnetic fluctuations are relatively weak, $\delta B_\perp/B_0 \ll 1$. In this case, we approximate:
\begin{eqnarray}
\Delta \ln(\gamma)\sim 2\phi \beta \frac{\delta B_\perp}{B_0}. 
\end{eqnarray}
In strong magnetically dominated  Alfv\'enic turbulence, the electric and magnetic fluctuations are nearly in equipartition, $E_\perp\sim \delta B_\perp$ \cite[e.g.,][]{tenbarge2021,chernoglazov2021,nattila2020,vega2022a}. Therefore, at small field-parallel scales, $l\ll L_{\|,0}$, the plasma velocity fluctuations can be evaluated as 
\begin{eqnarray}
\beta(l)\sim\frac{\delta B_\perp(l)}{B_0}\sim \frac{\delta B_0}{B_0}\left(\frac{l}{L_{\|, 0}}\right)^{1/2},   
\end{eqnarray}
where we used the field-parallel scaling of magnetic fluctuations characteristic of Alfv\'enic turbulence; see our discussion in the introduction. We then estimate the contribution of an eddy of scale $l$ to the energy gain as\footnote{As a consistency check, we verify that as a particle gets accelerated by a turbulent eddy, its drift in the field-perpendicular direction does not exceed the eddy field-perpendicular size~$\lambda$.  One can check that due to the curvature drift, a particle propagating through such an eddy, gets displaced in the field-perpendicular direction by a distance 
\begin{eqnarray}
\Delta R_\perp \sim \gamma \rho_0 \left(\frac{\delta B_0}{B_0} \right)\left(\frac{\lambda}{L_{\perp,0}} \right)^{1/3}.     
\end{eqnarray}
In Section~\ref{gamma_crit}: Eq.~(\ref{C4}), we demonstrate that the typical perpendicular gyroradius of such a particle is 
\begin{eqnarray}
\rho_\perp \sim \rho_0 \left(\frac{\delta B_0}{B_0} \right)^{3/2}\left(\frac{\rho_0}{L_{\perp,0}} \right)^{1/2}\gamma^{3/2},   
\end{eqnarray}
where $\rho_0=c/\Omega_s$ and $\Omega_s=q_s B/(m_s c)$ is the particle's cyclotron frequency. 
It is easy to see that the condition $\Delta R_\perp<\lambda$ is equivalent to the condition $\rho_\perp<\lambda$. Since particles can only be accelerated by eddies that are larger than their gyroradii, the condition $\Delta R_\perp<\lambda$ is always satisfied. }
\begin{eqnarray}
\label{en_gain}
\Delta \ln(\gamma)\sim 2\phi \left(\frac{\delta B_\perp(l)}{B_0}\right)^2\sim 2\phi \left(\frac{\delta B_0}{B_0}\right)^2 \frac{l}{L_{\|, 0}}.  
\end{eqnarray}

We see that the acceleration rate is sensitive to the relative strength of the turbulent fluctuations; it rapidly decreases as $\delta B_0/B_0$ decreases. We also notice that larger-scale fluctuations provide stronger acceleration, suggesting that the process is dominated by the particle interactions with the largest, outer-scale turbulent eddies,~$l\sim L_\|$. If we allow $\beta$ to attain both positive and negative values with certain probabilities, Eq.~(\ref{en_gain}) would lead to a logarithmic random walk. It would result in log-normal energy distributions of accelerated particles. The typical value of $\Delta \ln(\gamma)$ will then be evaluated as
\begin{eqnarray}
\label{en_gain2}
 \Delta \ln(\gamma)\sim 2\phi \left(\frac{\delta B_0}{B_0}\right)^2 \sqrt{N_0},  
\end{eqnarray}
where $N_0$ is the number of large-scale crossing times.

Formulae (\ref{en_gain}) and (\ref{en_gain2}) are our main result for the curvature acceleration in magnetically dominated Alfv\'enic turbulence. They have several important consequences. First, in contrast with the linear acceleration provided by a parallel electric field, the curvature acceleration is exponentially fast. Given long enough acceleration time (say, running time of numerical simulations), it would dominate over the energization provided by the parallel electric field. Second, in the limit of a strong guide field, $\delta B_0 \ll B_0$, the electron magnetic moment is conserved even when the electron interacts with intense intermittent structures. Therefore, as the electron accelerates, its field-perpendicular momentum does not significantly change, while the field-parallel momentum increases.  

We, therefore, propose that in magnetically dominated strong-guide-field Alfv\'enic turbulence, the acceleration is provided by curvature drifts, particles are accelerated along the magnetic field lines, and they attain log-normal energy distributions. Due to the quadratic dependence of the acceleration rate on the intensity of turbulent fluctuations, the acceleration time increases significantly when $\delta B_0/B_0$ decreases. These results are consistent with the numerical observations in \cite[][]{vega2024}. 
In Section~\ref{mirrors}, we will discuss the mirror acceleration mechanism. However, before that, we would like to present some relevant results on magnetic moment conservation in Alfv\'enic turbulence.

\subsection{Magnetic moment conservation in a turbulent magnetic field}
\label{gamma_crit}
In a collisionless plasma where fluctuations of the electric and magnetic fields are relatively small in comparison with $B_0$, a charged particle preserves the first adiabatic invariant, the magnetic moment. The complete expression for the magnetic moment  includes not only field-perpendicular but also field-parallel particle momentum \cite[e.g.,][]{northrop1963,littlejohn1983,littlejohn1984,egedal2008}. This is especially relevant for the case when the field-parallel momentum of a particle is much larger than its field-perpendicular momentum, $p_\perp/p_\|\ll 1$, which we consider here. In this limit, the leading order expression for the magnetic moment is $\mu^{(0)}=p_\perp^2/(2m_sB_0)$. The next-order term becomes relevant when 
$p_\perp \sim p_\|^2c/(q_sB_0R_c)$. 
We propose that as the relative value of the field-perpendicular momentum decreases, it cannot become smaller than the value dictated by balancing the leading and sub-leading terms in the expression for the magnetic moment. Therefore, in a curved magnetic field, as the field-parallel momentum increases during particle acceleration, so should the field-perpendicular momentum. For an ultrarelativistic particle, we may then obtain the typical value attained by the particle's perpendicular momentum as its parallel momentum increases:
\begin{eqnarray}
\label{C1}
p_\perp \sim p_\|^2c/(q_sB_0R_c)\sim p_\|\gamma\rho_0/R_c.   
\end{eqnarray}
Here $\rho_0=c/\Omega_s$ and $\Omega_s=q_sB/(m_s c)$ is the particle non-relativistic cyclotron frequency, $q_s$ its charge, and $m_s$ its rest mass. 
Expressing the particle gyroradius as $\rho_\perp = \gamma\rho_0\sin\theta$,  we rewrite formula (\ref{C1}) for a small pitch angle: 
\begin{eqnarray}
\label{C2}
\rho_\perp\sim \rho^2_0\gamma^2/R_c.   
\end{eqnarray}
If the particle gyroradus, $\rho_\perp$, is smaller than the inner scale of Alfv\'enic turbulence, $d_{rel}$, the largest curvature of the magnetic-field lines is provided by the smallest turbulent eddies, that is, the eddies with scales $d_{rel}$. The curvature radius can then be evaluated as in formula~(\ref{R_c}), 
\begin{eqnarray}
R_c\sim d_{rel} \left(\frac{B_0}{\delta B_\perp(d_{rel})}\right)^2\sim L_{\perp,0}\left( \frac{d_{rel}}{L_{\perp,0}}\right)^{1/3}\left( \frac{B_0}{\delta B_0}\right)^2. \quad \quad  
\end{eqnarray}
 Here, for simplicity, we assumed the \citet{goldreich_toward_1995} scaling of the magnetic fluctuations, $\alpha=1/3$. Substituting this expression into Equation~(\ref{C2}), we derive the scaling of the typical particle gyroradius with the energy:
\begin{eqnarray}
\label{rho_perp_small}
\rho_\perp\sim \rho_0 \left(\frac{\rho_0}{L_{\perp,0}} \right) \left( \frac{\delta B_0}{B_0} \right)^2 \left(\frac{L_{\perp,0}}{d_{rel}} \right)^{1/3}\gamma^2.
\end{eqnarray}
This gyroradius becomes comparable to the inner scale of turbulence, $d_{rel}$, when the particle energy reaches the {critical value}:
\begin{eqnarray}
\gamma_c=\frac{B_0}{\delta B_0}\frac{d_{rel}}{\rho_0}\left(\frac{L_{\perp,0}}{d_{rel}}\right)^{1/3}.   
\end{eqnarray}

For energies larger than the critical energy, the curvature of the magnetic field lines guiding the particle motion, is provided by the eddies comparable to the particle gyroradius, $\lambda\sim \rho_\perp$. In this case, the field curvature radius needs to be estimated differently, 
\begin{eqnarray}
\label{C3}
R_c\sim L_{\perp,0}\left( \frac{\rho_\perp}{L_{\perp,0}}\right)^{1/3}\left( \frac{B_0}{\delta B_0}\right)^2.    
\end{eqnarray}
Substituting Eq.~(\ref{C3}) into Eq.~(\ref{C2}), we derive the scaling of the particle gyroradius with the Lorenz factor,
\begin{eqnarray}
\label{C4}
\rho_\perp \sim \rho_0 \left(\frac{\delta B_0}{B_0} \right)^{3/2}\left(\frac{\rho_0}{L_{\perp,0}} \right)^{1/2}\gamma^{3/2}.   
\end{eqnarray}
This scaling holds only when $\rho_\perp$ is larger than the smallest scale of Alfv\'enic turbulence, $d_{rel}$. 

Interestingly, in this limit we can give an alternative derivation of formula~(\ref{C1}), which is more suitable for our analysis of magnetic turbulence.  In an Alfv\'enic turbulent eddy with the field-perpendicular and field-parallel scales $\lambda$ and $l$, the directions of magnetic field lines are known with the angular uncertainty of $\theta_\lambda \sim \lambda/l$. Therefore, a particle with a gyroradius $\rho_\perp\sim \lambda$ cannot maintain a pitch angle smaller than $\theta_\lambda$. Expressing the (small) particle pitch angle as $p_\perp/p_\|$, we write this condition as 
\begin{eqnarray}
\label{pitch_uncertainty}
\frac{p_\perp}{p_\|}\sim \frac{\lambda}{l}.    
\end{eqnarray}
The magnetic-line curvature associated with such an eddy can be evaluated as $R_c\sim l^2/\lambda$, and the scale of the eddy guiding the particle motion as $\lambda\sim \rho_\perp=\gamma \rho_0 \sin\theta$. If one expresses $\lambda$ and $l$ through $\rho_\perp$ and $R_c$, one can verify that Eq.~(\ref{pitch_uncertainty}) becomes equivalent to Eq.~(\ref{C1}).

It is also easy to see that the condition~(\ref{pitch_uncertainty}) or~(\ref{C4}) is the resonance condition between a particle and a turbulent eddy. When a particle's energy reaches the critical value $\gamma_c$, its gyroradius $\gamma\rho_0$ becomes comparable to the field-parallel size of the smallest Alfv\'enic eddy. The particle's gyrofrequency also becomes comparable to the turnover rate of the smallest eddy of scale $d_{rel}$ at this energy. This interaction leads to the scattering of the particle's pitch-angle. However, as the pitch angle increases and moves away from the resonance condition (\ref{C4}), the scattering becomes less effective. The curvature acceleration, on the other hand, tends to increase $\gamma$ and decrease the pitch angle. Therefore, it is reasonable to expect that at $\gamma>\gamma_c$, the resulting pitch-angle will be governed by the ``critical balance" condition~(\ref{C4}).

\subsection{Mirror acceleration}
\label{mirrors}
\begin{figure}[tb]
\centering
\includegraphics[width=0.8\columnwidth]{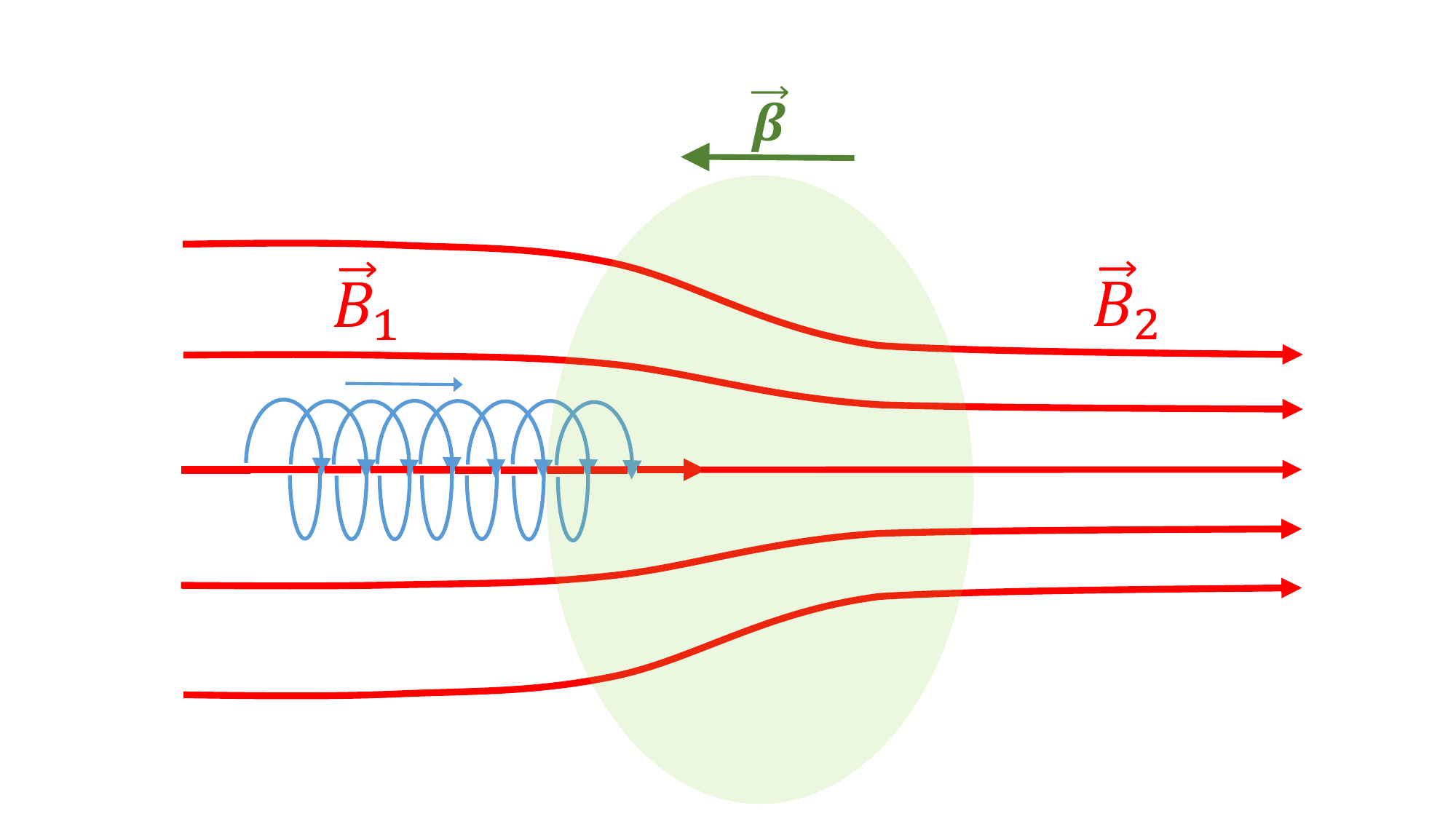}
\includegraphics[width=0.8\columnwidth]{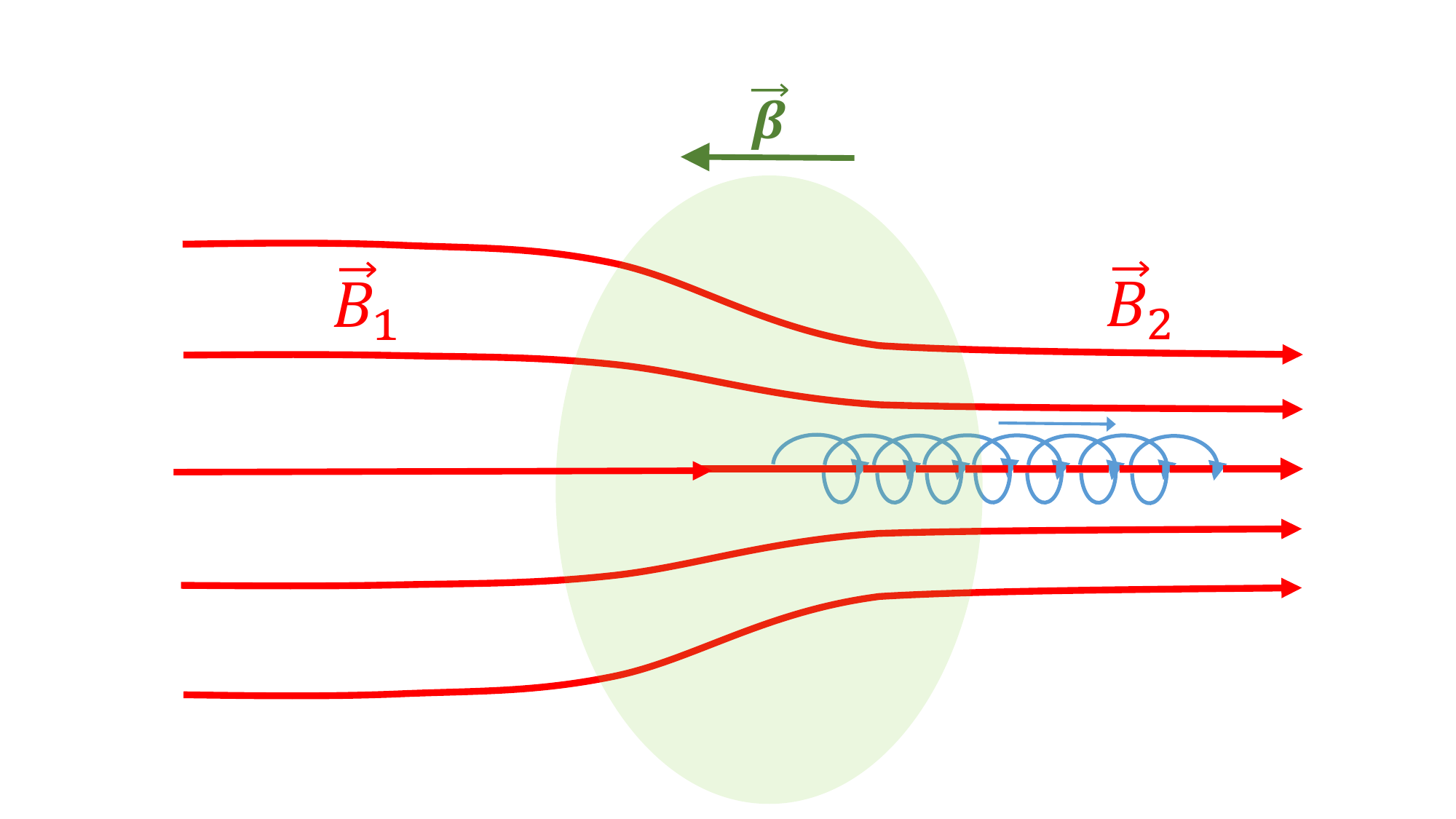}
\caption{Sketch of an electron-mirror interaction for {\it small} pitch angles. For simplicity, velocity ${\bm \beta}$ is directed along the magnetic mirror axis.}
\label{mirror_collision1}
\end{figure}

Curvature acceleration discussed in Section~\ref{curve_section} does not significantly depend on the pitch angle in that it remains efficient even when the pitch angle is small. In contrast, the mirror acceleration depends crucially on the value of the pitch angle. Fig.~\ref{mirror_collision1} shows a head-on interaction of an electron with a magnetic mirror when the electron pitch angle is small, $\sin^2\theta< 1-(\Delta B/B_2)$. Here, $\Delta B=B_2-B_1>0$ is the variation of the magnetic field strength, as shown in the figure.
In this case, the electron is not reflected by the mirror but rather propagates through the mirror throat. For simplicity, assume that the electron is ultra-relativistic and the mirror is moving at a relatively lower speed. When the electron passes through the mirror, its energy increases approximately as 
\begin{eqnarray}
\label{mirror_small}
\Delta \ln(\gamma) \sim \frac{1}{2} \beta \,\left(\frac{\Delta B/B_2}{1-\Delta B/B_2}\right){\sin^2\theta},   
\end{eqnarray}
where we assumed $\sin^2\theta \ll 1$. We see that for small pitch angles, the mirror acceleration given by Eq.~(\ref{mirror_small}) is much less efficient than the acceleration by curvature. Due to the conservation of magnetic moment in Alfv\'enic turbulence with a strong guide field, $\delta B_0\ll B_0$, particle acceleration decreases the pitch angle even further, $\sin^2\theta\sim 1/\gamma^2$. Therefore, the mirror acceleration becomes even less relevant. 

The situation changes drastically in the case of a moderate guide field, $B_0\sim \delta B_0$. In this case, collisions with intermittent structures with thickness $d_{rel}$ and magnetic-field variations $\delta {\bm B}_0$ provide significant pitch-angle scattering to particles whose gyroradius exceeds $d_{rel}$. When a particle is accelerated in a curved magnetic field, its gyroradius depends on its energy $\gamma$. As demonstrated in Section~\ref{gamma_crit}, in Alfv\'enic turbulence the particle's gyroradius exceeds $d_{rel}$ when its energy exceeds the  critical value, $\gamma \geq \gamma_c$ given by
\begin{eqnarray}
\label{gamma_critical}
\gamma_c=\frac{B_0}{\delta B_0}\frac{d_{rel}}{\rho_0}\left(\frac{L_{\perp,0}}{d_{rel}}\right)^{1/3}.   
\end{eqnarray}   
In this formula, $\rho_0=c/\Omega_s$ and $\Omega_s=q_s B/(m_s c)$ is the particle cyclotron frequency.

\begin{figure}[tb]
\centering
\includegraphics[width=0.8\columnwidth]{collide2.pdf}
\includegraphics[width=0.8\columnwidth]{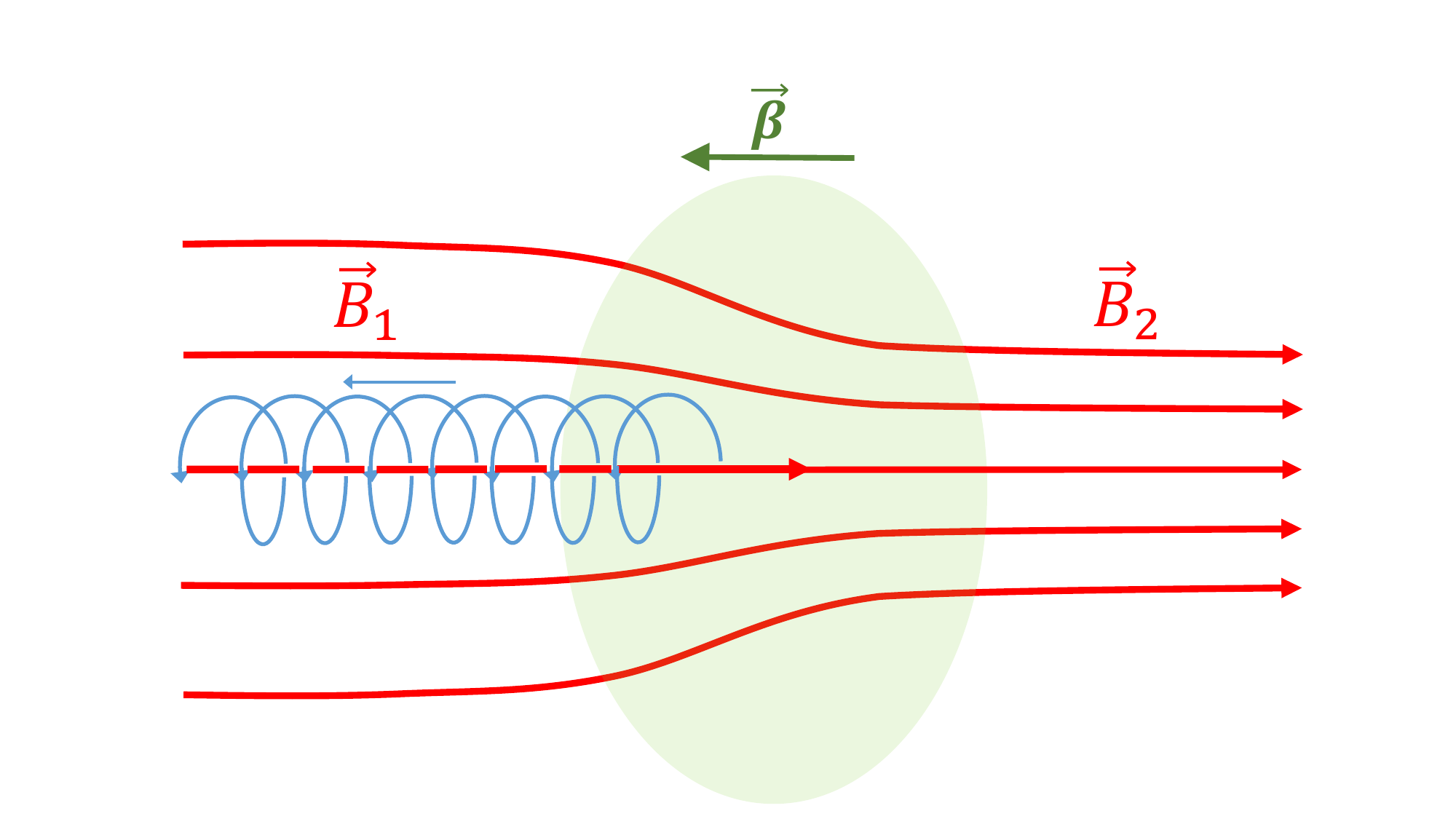}
\caption{Sketch of an electron-mirror interaction for {\it large} pitch angles. For simplicity, velocity ${\bm \beta}$ is directed along the magnetic mirror axis.}
\label{mirror_collision2}
\end{figure}

At such energies, and when $\delta B_0/B_0\sim 1$, energetic particles can be scattered to large pitch angles, $\sin^2\theta> 1-\Delta B/B_2$, in which case they may be efficiently accelerated by mirrors. Such a situation is shown in Fig.~\ref{mirror_collision2}.
In such an interaction, a particle gets reflected from the mirror, and its energy increases according to
\begin{eqnarray}
\label{mirror_large}
\Delta \ln(\gamma)\sim 2\beta \cos\theta.    
\end{eqnarray}
This energy gain is larger than that associated with a small pitch angle, Eq.~(\ref{mirror_small}). It is comparable to the curvature acceleration at a similar guide-field strength. Therefore, in the case of a moderate guide field, the population of particles with large pitch angles satisfying $\sin^2\theta> 1-(\Delta B/B_2)$, is accelerated by both curvature and mirror effects. As a result, such particles gain energy at a higher exponential rate than the population with small pitch angles, $\sin^2\theta < 1-(\Delta B/B_2)$. 

This consideration is broadly consistent with the model of particle acceleration proposed in \cite[][]{vega2022a}. In this model, particles are accelerated fast in the phase-space region defined by the pitch angle condition, $\sin^2\theta > 1-(\Delta B/B_2)$. Due to pitch-angle scattering, particles leak from this region when their pitch angles become smaller. At smaller pitch angles, the acceleration is exponentially weaker and it is neglected. Such a ``phase-space leaky box" leads to power-law energy distributions of accelerated particles. The power law exponent depends on the rate of particle leakage, which in turn, is a function of the turbulence intensity, $\Delta B/B_2$.

It is also worth noting that in Alfv\'enic turbulence, mirror structures are generally spatially separated from the regions of large curvature, and, therefore, may be considered as complementary effects.  Indeed, phenomenological arguments and numerical simulations suggest that magnetic strength and magnetic curvature are anti-correlated in Alfv\'enic turbulence \cite[e.g.,][]{schekochihin2004,kempski2023}. Moreover, magnetic mirrors are also separated from the intermittent structures that provide pitch-angle scattering. Indeed, strong magnetic shears in Afv\'enic turbulence are typically associated with rotations of magnetic field direction rather than with variations of magnetic field strength \cite[e.g.,][]{zhdankin2012a,zhdankin2012b}.

\section{Numerical Illustration}
\label{examples}
To illustrate the discussion in the previous section, we show the results from two 2.5D numerical simulations of decaying turbulence in a pair plasma. The simulations were run with fully relativistic particle-in-cell code VPIC \cite[][]{bowers2008}. A 2.5D simulation can be seen as a 3D simulation with continuous translational symmetry along one direction (the $z$ direction in our case), so only one two-dimensional cut needs to be simulated.
\begin{table}[tb]
\vskip5mm
\centering
\begin{tabular}{c c c c c c} 
\hline
{Run} & {Size} ($d_e^2$) & \# of cells & $\omega_{pe}\delta t$ & $B_0/\delta B_0$ & $\tilde{\sigma}$\\
\hline
I & $2000^2$ & $16640^2$ & $2.1\times10^{-2}$ & 1 & 40\\ 
II & $1600^2$ & $23552^2$ & $1.2\times10^{-2}$ & 10 & 40\\ 
\hline
\end{tabular}
\caption{Parameters of the runs. Here, $d_e$ is the nonrelativistic electron inertial scale. The table also shows the initial value of~$B_0/\delta B_0$ and the initial value of the magnetization ${\tilde \sigma}$ defined by formula~\eqref{sigma}. Some results of these simulations we also analyzed in \cite[][]{vega2022a,vega2024}, where more specific details of the initial setup can be found.}
\label{table}
\end{table}

Both simulation domains were double periodic $L\times L$ squares with 100 particles per cell per species and had a uniform magnetic guide-field $\bm{B}_0=B_0\hat{\bm{z}}$. Turbulence was initialized with randomly phased magnetic perturbations of the Alfv\'enic type
\begin{eqnarray}
\delta{\bm B}(\mathbf{x})=\sum_{\mathbf{k}}\delta B_\mathbf{k}\hat{\xi}_\mathbf{k}\cos(\mathbf{k}\cdot\mathbf{x}+\phi_\mathbf{k}),
\end{eqnarray}
where the unit polarization vectors are normal to the background field, $\hat{\xi}_\mathbf{k}=\mathbf{k}\times {\bm B}_0/|\mathbf{k}\times{\bm B}_0|$. The wave vectors of the modes are given by $\mathbf{k}=\{2\pi n_x/L,{2}\pi n_y/L\}$, where $n_x,n_y=1,...,8$. All modes have the same amplitudes $\delta B_{\mathbf{k}}$. The rms value of the perturbations is given by $\delta B_0={\langle |\delta{\bm B}(\mathbf{x})|^2 \rangle^{1/2} }$. The initial temperature of the particles was chosen to be $\theta_0\equiv k_BT_0/m_ec^2= 0.3$, which corresponds to the initial enthalpy of $w_0\approx 1.88$. In the figures below, time is normalized to the large-scale light crossing time $L_{\perp,0}/c$, where $c$ is the speed of light and $L_{\perp,0}=L/8$ is the outer scale of turbulence. Table \ref{table} summarizes the parameters of the runs.

\begin{figure*}[tb]
\centering
\includegraphics[width=0.97\columnwidth]{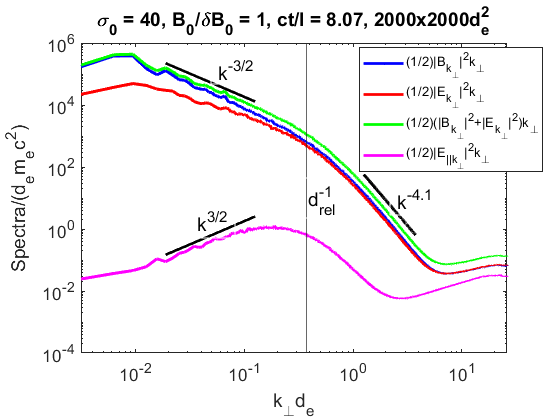}
\includegraphics[width=0.97\columnwidth]{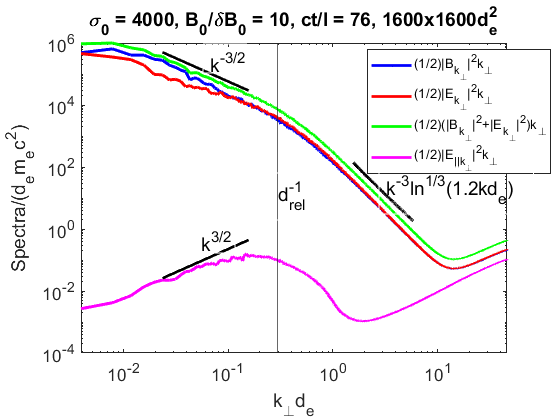}
\caption{Electric and magnetic energy spectra for simulations with $B_0/\delta B_0=1$ (left panel) and $B_0/\delta B_0=10$ (right panel). Note rather weak parallel electric field fluctuations. The slopes indicated by the solid black lines are given for the reader's orientation. They are chosen to be consistent with the discussions given in \cite[][]{vega2022a,vega2024}, where some results of these numerical simulations were also analyzed. }
\label{EM_spectra}
\end{figure*}

\begin{figure*}[tb]
\centering
\includegraphics[width=0.97\columnwidth]{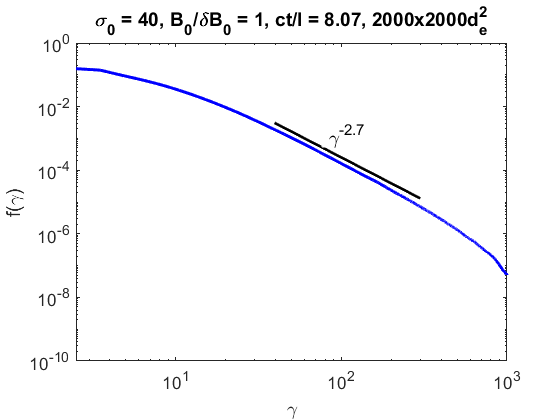}
\includegraphics[width=0.97\columnwidth]{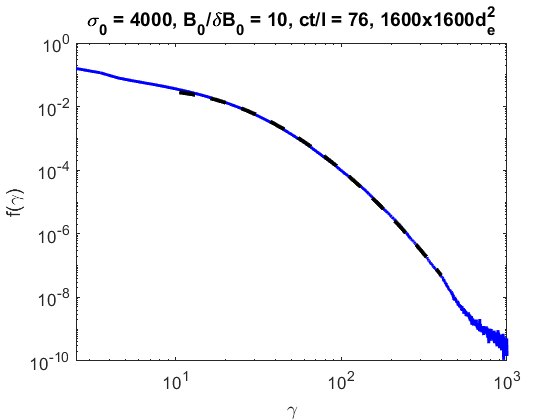}
\caption{Particle energy probability density function for simulations with $B_0/\delta B_0=1$ (left panel) and $B_0/\delta B_0=10$ (right panel). In the case of a moderate guide field, the energy distribution function has a power-law tail. In the case of a strong guide field, the high-energy tail is well approximated by a log-normal function $f(\gamma)\propto \gamma^{-1}\exp[-(\ln\gamma-\mu)^2/2\sigma_s^2]$, with $\mu\approx 2.6$ and $\sigma_s^2\approx 0.61$. This function is shown by the dashed line. Note the significantly longer time required to accelerate the particles in the case of a strong guide field. }
\label{particle_pdf}
\end{figure*}

\begin{figure*}[tb]
\centering
\includegraphics[width=0.97\columnwidth]{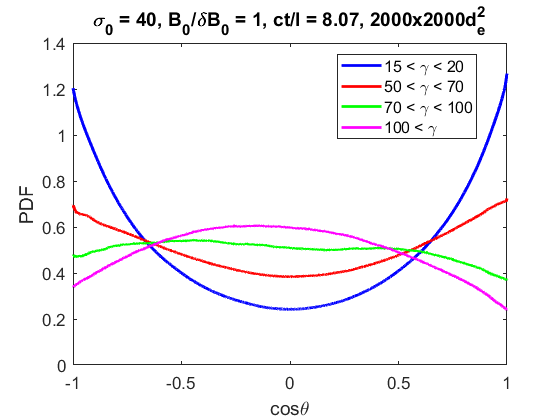}
\includegraphics[width=0.97\columnwidth]{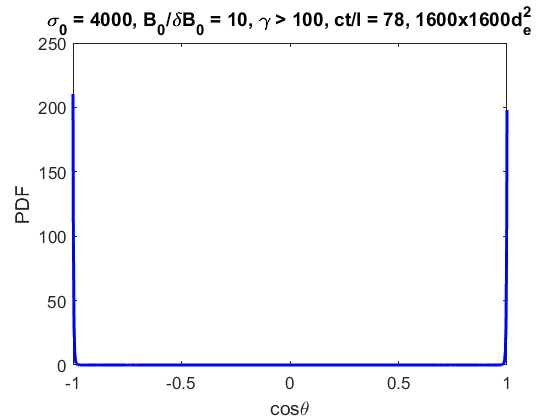}
\caption{Pitch-angle distribution of ultrarelativistic particles ($\gamma>100$) for simulations with a moderate guide field, $B_0/\delta B_0=1$ (left panel) and a strong guide field, $B_0/\delta B_0=10$ (right panel). In the case of a strong guide field, the pitch angles are extremely small. In the case of a moderate guide field, the pitch angles of accelerated particles crucially depend on the particle's energy. For energies smaller than the critical energy, $\gamma \ll \gamma_c$ (in this run, $\gamma_c\approx 70$), the particles are accelerated mostly along the background magnetic field. Above the critical energy, $\gamma\gtrsim\gamma_c$, particles with larger pitch angles are accelerated more efficiently.}
\label{particle_angles}
\end{figure*}

Both runs are very similar in the way the Alfv\'enic turbulence is initialized. The only significant difference is the strength of the applied guide field, ${\bm B}_0$. In both cases, the most efficient phase of particle heating continued until the growing energy of particles approaches an approximate equipartition with the decaying energy of electromagnetic fluctuations. This happened when about half of the initial magnetic energy was transferred to particles, so their energies reached $\langle\gamma \rangle\sim 10$ for each species. The electromagnetic spectra of turbulence and the energy distribution functions of the electrons are shown at these moments in Figures~\ref{EM_spectra} and~\ref{particle_pdf}. Figure~\ref{EM_spectra} shows that the spectra of electromagnetic fluctuations in the Alfv\'enic interval, $k_\perp d_{rel}\lesssim 1$, are similar in both runs. 

Note, however, that it took a significantly longer time to energize the particles in the case of a strong guide field than in the case of a moderate guide field, in qualitative agreement with our discussion. Moreover, the resulting non-thermal tails ($\gamma > 10$) of the energy distribution functions are drastically different in the two cases.  In the case of a strong guide field, the distribution is well approximated by a log-normal function. In the case of a moderate guide field, the distribution is close to a power law. This also qualitatively agrees with our modeling. 

Figure~\ref{particle_angles} shows the distributions of pitch angles of accelerated particles. In the case of a strong guide field, the pitch angles are extremely small. This is consistent with our formula for the critical energy, Eq.~(\ref{gamma_critical}), that gives for this case $\gamma_c\approx 8000$. This means that even the most energetic particles generated in our run, with the energies $\gamma\sim 1000$, will have gyroradii much smaller than~$d_{rel}$. The process of curvature acceleration is described by Eq.~(\ref{en_gain2}), leading to a log-normal particle energy distribution.  

The situation is fundamentally different in the case of a moderate guide field. Here, the critical energy estimate provided by Eq.~(\ref{gamma_critical}), gives $\gamma_c\approx 70$. Figure~\ref{particle_angles}, left panel, indeed shows that particles with the smaller energies, $\gamma \ll 70$, are accelerated mostly along the magnetic field lines. However, particles with higher energies $\gamma \gtrsim 70$ are more efficiently accelerated when they have large pitch angles. This is consistent with our picture of particle acceleration suggesting that in the case of a moderate guide field, $B_0\sim \delta B_0$,  particles with $\gamma\gtrsim \gamma_c$ experience strong pitch-angle scattering and efficient acceleration by magnetic mirrors.

\section{Discussion}
We have proposed a phenomenological description of non-thermal relativistic particle acceleration in a magnetically dominated Alfv\'enic turbulence. We considered the setting where turbulence is excited by initial field perturbations with strong magnetization, ${\tilde \sigma_0 \gg 1}$. When such perturbations decay, they drive turbulence and energize the plasma. As a result, plasma is self-consistently heated to ultrarelativistic temperatures, while simultaneously, the developed Alfv\'enic turbulence leads to non-thermal particle acceleration.   
We argue that in the considered limit of strong magnetization, the process of particle acceleration is universal in that it depends only on the relative strength of the imposed guide magnetic field and the resulting turbulent fluctuations. The process of acceleration is governed by the conservation of magnetic moment. In the case of a strong guide field, $B_0\gg \delta B_0$, the particle's magnetic moment is conserved and the acceleration is provided by magnetic curvature drifts.  The curvature acceleration energizes particles in the direction parallel to the magnetic field lines, resulting in a log-normal tail of particle energy distribution function. The situation is qualitatively different in a setting when plasma is immersed in a moderate guide field, $B_0 \lesssim \delta B_0$. In this case, as the gyroradius of an energetic particle exceeds the inner scale of turbulence (in our case, $d_{rel}$), interactions with intense turbulent structures like current sheets can break the particle's magnetic moment. Magnetic mirror effects become important at such energies, resulting in power-law energy distributions of accelerated particles. The proposed phenomenological picture is consistent with available numerical simulations, as illustrated in the examples discussed in section~\ref{examples}. \\

This material is based upon work supported by the U.S. Department of Energy, Office of Science, Office of Fusion Energy Sciences under award number DE-SC0024362. The work of CV and SB was also partly supported by the NSF grant PHY-2010098 and the Wisconsin Plasma Physics Laboratory (US Department of Energy Grant DE-SC0018266). VR was also partly supported by NASA grant 80NSSC21K1692. Computational resources were provided by the Texas Advanced Computing  Center (TACC) at the University of Texas at Austin and by the NASA High-End Computing (HEC) Program through the NASA Advanced Supercomputing (NAS) Division at Ames Research Center. This work also used Bridges-2 at Pittsburgh Supercomputing Center. TACC and Bridges-2 access was provided by allocation TG-ATM180015 from the Advanced Cyberinfrastructure Coordination Ecosystem: Services \& Support (ACCESS) program, which is supported by National Science Foundation grants \#2138259, \#2138286, \#2138307, \#2137603, and \#2138296.\\

%



\appendix
\section{Polarization of the electric field in Alfv\'enic turbulence}
\label{e_polarization}
Here we follow the discussion of the linear Alfv\'en waves in a collisionless, relativistically hot magnetically dominated pair plasma presented in \cite[][]{vega2024}, see also \cite[][]{godfrey1975,gedalin1998}. Assume that $\Omega^2_{ce}\gg \omega^2_{pe}\gg \omega^2$, where $\Omega_{ce}$ and $\omega_{pe}$ are the electron cyclotron and plasma frequencies, correspondingly. Conventionally, the background magnetic field ${\bm B}_0$ is in the vertical $z$-direction, while the wave vector has the coordinates ${\bm k}=(k_\perp, 0, k_z)$.  Under these assumptions, the electric polarization of the Alfv\'en mode satisfies:
\begin{eqnarray}
\label{A1}
{\left(\begin{array}{ccc}
k_{z}^{2}-\frac{\omega^{2}}{c^{2}} & 0 & -k_{z}k_{\perp}\\
0 & k^{2}-\frac{\omega^{2}}{c^{2}} & 0\\
-k_{z}k_{\perp} & 0 & k_{\perp}^{2}-\frac{\omega^{2}}{c^{2}}P
\end{array}\right)}\left(\begin{array}{c}
E_{x}\\
E_{y}\\
E_{z}
\end{array}\right)=0.
\end{eqnarray}
We assume that the plasma particle distribution function is {\it one-dimensional}, $p_z\gg p_\perp$, and {\it ultra-relativistic}. For the field-parallel and field-perpendicular wave numbers satisfying $k^2_z\ll k_\perp^2$ and $k_\perp^2d_{rel}^2\lesssim 1$, one then estimates
\begin{eqnarray}
\label{A2}
P\approx -\frac{2w_0\omega_{pe}^{{2}}}{k_z^2 c^2}.
\end{eqnarray}
Here, $w_0$ is the specific enthalpy associated with the electron distribution, $d^2_{rel}=w_0 c^2/(2\omega_{pe}^2)$ the corresponding relativistic electron inertial scale, and $\omega^2\approx k_z^2 c^2$ the Alfv\'en wave frequency in a magnetically dominated plasma.   
From Equations~(\ref{A1}) and (\ref{A2}) we derive that $E_y=0$, and:
\begin{eqnarray}
\label{A3}
\frac{E_\|}{E_\perp}\equiv \frac{E_z}{E_x}\approx\frac{1}{w_0^2}k_zk_\perp d^2_{rel}.    
\end{eqnarray}
{~}\\

\bibliography{references}{}
\bibliographystyle{aasjournal}



\end{document}